\documentclass[doublecol]{epl2}
\usepackage{amsfonts}
\usepackage{amsmath}
\usepackage{hyperref}
\usepackage{subcaption}
\usepackage{graphicx}
\usepackage{slashed}
\newcommand\correspondingauthor{\thanks{E-mail: \href{mailto:aobispo@utp.edu.pe}{aobispo@utp.edu.pe} (corresponding author)}}
\hypersetup{colorlinks=true,linkcolor=blue,citecolor=blue,filecolor=blue,urlcolor=blue,breaklinks=true}

\title{Remarks on the ($2+1$)-dimensional Duffin-Kemmer-Petiau oscillator in an external magnetic field}
\shorttitle{Remarks on the ($2+1$)-dimensional DKPO in an external magnetic field} %Insert here a short version of the title if it exceeds 70 characters

\author{Andr\'{e}s G. Jir\'{o}n\inst{1}\and Luis B. Castro\inst{2}\and Angel E. Obispo\inst{1,3}\!\!\!\correspondingauthor \and Antonio S. de Castro\inst{4}}
\shortauthor{Andrés G. Jirón\etal}

\institute{
  \inst{1} Universidad Tecnológica del Perú (UTP), Los Olivos, Lima, Perú\\
  \inst{2} Universidade Federal do Maranhão (UFMA), S\~{a}o Luís, MA, Brazil\\
  \inst{3} Universidad Privada del Norte (UPN), Los Olivos, Lima, Perú\\
  \inst{4} Universidade Estadual Paulista (UNESP), Guaratinguetá, SP, Brazil}
\abstract{This work re-examines the issue of spin-$1$ particles in a ($2+1$)-dimensional Duffin-Kemmer-Petiau oscillator (DKPO) in the presence of an external magnetic field. By following the appropriate procedure for the spin-$1$ sector of the Duffin-Kemmer-Petiau (DKP) theory, the previously used $6\times 6$ representation in the literature is shown to be reducible to a $3\times 3$ irreducible representation. This approach enabled us to find new aspects of the results recently disseminated in various studies, as well as other considerations overlooked and requiring revision. Finally, we present some applications of two-dimensional DKP theory in condensed matter systems, particularly in Lieb lattices.}
\begin{document}

\maketitle

\section{INTRODUCTION}
\label{intro}
The Duffin-Kemmer-Petiau equation (DKP) \cite{Petiau1936,Kemmer1938,PR54:1114:1938,Kemmer1939} is a first-order covariant wave equation that describes spin-$0$ and spin-$1$ fields or particles with a unique algebraic structure not expressible by the traditional Klein-Gordon (KG) and Proca theories \cite{PRD15:1518:1977,JPA12:665:1979}. The DKP equation is similar to the Dirac equation, but with beta-matrices replacing gamma-matrices and satisfying the DKP algebra \cite{PR54:1114:1938,PTP10:589:1953}. In $(3+1)$-dimensional space-time, the DKP algebra generates a set of $126$ linearly independent matrices \cite{CORSON1953} with irreducible representations consisting of a trivial representation, a five-dimensional representation for the scalar particles (spin-0 sector), and a ten-dimensional representation for the vectorial particles (spin-1 sector). Thanks to its rich variety of interactions, the DKP theory has been applied to solve various problems, including scattering of mesons by nuclei \cite{NPA585:311:1995}, $\alpha$-nucleus scattering, deuteron-nucleus elastic scattering \cite{NPA589:307:1995,PRC40:2181:1989}, the quark confinement problem of quantum chromodynamics \cite{EPJC10:71:1999}, and some other systems based on minimally coupled vector interactions \cite{PLA244:329:1998,PLA268:165:2000,PRA90:022101:2014}. Additionally, with respect to systems with non-minimal coupling, there has been increasing interest in the so-called DKP oscillator (DKPO) \cite{ZPC56:421:1992,JPA31:3867:1998,JPA43:495402:2010}.

The DKPO arises as a kind of tensor coupling with a linear potential that leads to the harmonic oscillator problem in the weak-coupling limit. Although the DKP equation and the DKPO are widely studied models in four-dimensional space, recently, a great deal of theoretical research has been done on these systems in $(2+1)$-dimensional space-time, which are motivated by their potential applications in other areas of physics, for instance, in condensed matter systems \cite{PRB92:235106:2015,PRB97:075135:2018,PRB102:207101:2020}. However, the DKP algebra in $(2+1)$ dimensions only generates a set of $35$ linearly independent matrices, leading to a trivial representation, a four-dimensional representation (spin-$0$ sector), and a twin three-dimensional representation (spin-$1$ sector). Interestingly, this twin reducible representation can be combined to extend the beta-matrices to a six-dimensional representation \cite{PTP10:589:1953,CORSON1953}. The DKPO for spin-$1$ bosons in an external magnetic field has been studied in this context \cite{EPJC72:2217:2012,CJP91:1:2013,ADHEP2014:185169:2014,EPJP132:186:2017,%
ChPB27:010301:2018,MPLA35:2050278:2020,PLA384:126706:2020,PLA433:128030:2022,chargui2023spin}. Particularly, in Ref. \cite{chargui2023spin} the authors demonstrate the emergence of a $6\times 6$ matrix representation resulting from the dimensional reduction of the DKPO in $(3+1)$-dimensional space-time using $10\times 10$ beta-matrices. Exact solutions for all physical components are computed in a simplified manner by decoupling the original system into two $3\times 3$ subproblems, without implying a reduction in the dimension of the beta-matrices. However, this system does not strictly correspond to a scenario in $(2+1)$-dimensional space-time, but represents a system in $(3+1)$-dimensions confined to the plane ($z=p_z=0$). Consequently, this approach does not allow for reducing the dimension of beta-matrices to a $3\times 3$ form, as this irreducible representation corresponds to an inherently 2D-dimensional system, as indicated by Corson \cite{CORSON1953}.

In this work, we revisit the problem of spin-$1$ particles in a $(2+1)$-dimensional Duffin-Kemmer-Petiau oscillator (DKPO) in the presence of an external magnetic field. We adopted the $6\times 6$ representation used in Ref.~\cite{PLA384:126706:2020} and we demonstrate its reduction to two equivalent $3\times 3$ representations, facilitating the derivation of the DKP spinor and energy spectrum in a more straightforward manner. Our solutions provide a corrected version of those presented in \cite{PLA384:126706:2020}, where some components are eliminated for simple mathematical convenience, generating incomplete equations with wrong relations between physical components, as already criticized in \cite{PLA433:128030:2022}. This wrong treatment is also found in Refs. \cite{EPJC72:2217:2012,CJP91:1:2013,ADHEP2014:185169:2014,EPJP132:186:2017,%
ChPB27:010301:2018,MPLA35:2050278:2020}.

This work is organized as follows. In Sec. 2, we discuss some aspects of the DKP equation in $(2+1)$-dimensional space-time. In Sec. 3, we concentrate our efforts in the interaction called DKP oscillator plus the space component of a four-vector. We focus on the case of vector bosons and obtain the equation of motion, energy spectrum and DKP spinor. In Sec. 4, we investigate two particular cases: DKPO without magnetic field (4.1) and DKP equation in presence of a uniform magnetic field (4.2). In Sec. 5, we present some applications of this 2D-DKP in condensed matter systems, specifically in computing certain electronic properties in Lieb lattice. Finally, in Sec. 6 we present our final remarks.

\section{The Duffin-Kemmer-Petiau equation}\label{sec:DKP}
\noindent The first-order Duffin-Kemmer-Petiau (DKP) equation for a free boson of rest mass $m$ is given by \cite{Kemmer1939} ($\hbar=c=1$)
\begin{equation}\label{ec-dkp}
(\beta^\mu p_\mu-m)\Psi=0,
\end{equation}
\noindent where the matrices $\beta^{\mu}$ satisfy the following algebra
\begin{equation}\label{algebradkp}
    \beta^\mu \beta^\nu \beta^\eta+\beta^\eta \beta^\nu \beta^\mu =g^{\mu \nu} \beta^\eta+g^{\eta \nu }\beta^\mu,
\end{equation}
\noindent and $g^{\mu \nu}$ is the Minkowski metric with the signature $(+,-,-,-)$. The conserved four-current is defined by
\begin{equation}\label{4current}
J^{\mu}=\frac{1}{2}\bar{\Psi}\beta^{\mu}\Psi\,,
\end{equation}
\noindent where the adjoint spinor $\bar{\Psi}$ is given by $\bar{\Psi}=\Psi^{\dagger}\eta^{0}$ with $\eta^{0}=2\beta^{0}\beta^{0}-1$ \cite{PRA90:022101:2014}. It is known that the time component of $J^{\mu}$ is not positive definite, but it may be interpreted as a charge density. The normalization condition for bound-state solutions is $\int d\tau J^{0}=\pm 1$,  where the plus (minus) sign must be used for a positive (negative) charge. In ($3+1$)-dimensional space-time, the algebra (\ref{algebradkp}) furnishes a set of $126$ independent matrices, which are part of three irreducible representations: $(i)$ a trivial representation (no physical meaning), $(ii)$ a five-dimensional representation (spin-$0$ sector) and $(iii)$ a ten-dimensional representation (spin-$1$ sector). From here on, we concentrate our attention on the spin-$1$ sector of the DKP theory.

To select the physical components of the DKP spinor for the spin-$1$ sector, we define the operators \cite{PTP10:589:1953,UMEZAWA1956}
\begin{equation}\label{r1}
    R^\mu \equiv (\beta^1)^2(\beta^2)^2(\beta^3)^2 (\beta^\mu \beta^0-g^{\mu 0}  ),
\end{equation}
\noindent and $R^{\mu\nu}=R^{\mu}\beta^{\nu}$, which satisfies $R^{\mu\nu}=-R^{\nu\mu}$. Using the projection operators, it is shown that all elements of the column matrix $R^\mu \Psi$ (physical components of the DKP spinor) obey the Proca equation,
\begin{equation}\label{eqpro1}
   \left( \partial^{\mu}\partial_{\mu}+m^{2}\right)R^\mu \Psi = 0 
\end{equation}
\begin{equation}\label{eqpro2}
   \partial_{\mu}R^\mu \Psi = 0 
\end{equation}
\noindent making explicitly clear that they select the spin-$1$ sector of the DKP theory \cite{UMEZAWA1956,PLA244:329:1998,PLA268:165:2000,PRA90:022101:2014}. From equation (\ref{eqpro2}), one can see that is possible to express one component in terms of the other three components. This fact means that only three physical components are linearly independent, which are related to the three degrees of freedom for a massive spin-$1$ particle.

\subsection{($2+1$)-dimensional DKP equation}

In ($2+1$) dimensions, the algebra (\ref{algebradkp}) generates a set of $35$ independent matrices \cite{CORSON1953} whose irreducible representations are: $(i)$ a trivial representation (no physical meaning), $(ii)$ a four-dimensional representation (spin-$0$ sector) and $(iii)$ a twin three-dimensional representation (spin-$1$ sector). It is worthwhile to mention that twin representations can be combined and they form a six-dimensional representation. From this fact, the authors in \cite{JMP60:082302:2019,PLA384:126706:2020,chargui2023spin,PLA433:128030:2022} claim that in ($2+1$) dimensions we should use a six-dimensional representation for the spin-$1$ sector of the DKP theory.

In ($2+1$) dimensions, the definition of the operator (\ref{r1}) becomes
\begin{equation}\label{r2}
    R^\mu \equiv (\beta^1)^2(\beta^2)^2 (\beta^\mu \beta^0-g^{\mu 0}  ),
\end{equation}
\noindent where $\mu$ takes the values from $0$ to $2$. The six-dimensional representation for the matrices $\beta^{\mu}$ used in \cite{JMP60:082302:2019,PLA384:126706:2020,PLA433:128030:2022} is given by
\begin{equation}\label{r6x6}
     \beta^{0} = \begin{pmatrix}
    \boldsymbol{0} & \rho_0  \\
 \rho_0& \boldsymbol{0} 
\end{pmatrix}, \hspace{0.2cm} \beta^j = \begin{pmatrix}
    \boldsymbol{0} & \rho_j \\
 -\rho_j^T& \boldsymbol{0}
\end{pmatrix}, \hspace{0.2cm} j=1,2  
\end{equation}
where $\boldsymbol{0}$ is a null matrix $3\times 3$, and
\begin{equation}\label{r6x61}
     \rho_{0} = \begin{pmatrix}
     -1 & 0& 0 \\
  0 & -1& 0\\
  0 & 0& 0
\end{pmatrix},\hspace{0.2cm} \rho_1 = \begin{pmatrix}
     0 & 0& 1\\
  0 & 0& 0\\
 0 & 1& 0
\end{pmatrix}, \hspace{0.2cm} \rho_2 = \begin{pmatrix}
     0 & 0& 0\\
  0 & 0& 1\\
 -1 & 0& 0
\end{pmatrix}.
\end{equation}
\section{DKP oscillator in a magnetic field}\label{section:dkpo_mf}

The introduction of the interaction called DKP oscillator plus the space component of a four-vector ($\vec{A}$) can be performed through substitution
\begin{equation}\label{aco_min}
\vec{p}\rightarrow \vec{p}-q\vec{A}-im\omega\eta^{0}\vec{r}\,,
\end{equation}
\noindent where $q$ is the charge of the vector boson and $\omega$ is the oscillator frequency. For time-independent interactions, one can write $\Psi(\vec{r},t)=\psi(\vec{r})\mathrm{exp}(-iEt)$, where $E$ is the energy of the vector boson, so that the time-independent DKP equation becomes
\begin{equation}\label{dkp_it}
    \left[-\beta^0 E+\vec{\beta}\cdot\left(\vec{p}-q\vec{A}-im\omega \eta^0 \vec{r} \right)+m \right]\psi=0\,.
\end{equation}
\noindent Considering $\vec{A}=\frac{B}{2}(-y,x)$, $\vec{r}=(x,y)$, and $\psi=(a_{1},a_{2},b,d_{1},d_{2},e)^{T}$ (following the notation of Ref. \cite{PLA384:126706:2020}), the equation (\ref{dkp_it}) decomposes into
\begin{equation}
m a_1 = -Ed_1- \pi^+_x e, \label{ec4}
\end{equation}
\begin{equation}
m a_2 = -Ed_2-\pi^+_y e, \label{ec5}
\end{equation}
\begin{equation}
m b  = -\pi^-_xd_2  + \pi^-_yd_1, \label{ec6}
\end{equation}
\begin{equation}
m d_1 =-Ea_1-\pi^+_y b, \label{ec7}
\end{equation}
\begin{equation}
m d_2 =-Ea_2+\pi^+_xb, \label{ec8}
\end{equation}
\begin{equation}
me  = \pi^-_y a_2 + \pi^-_xa_1.
\label{ec9}
\end{equation}
\noindent where $\tilde{\omega}=qB/(2m)$ and
\begin{equation}
    \pi^\pm _x = p_x \mp im\omega x + m\tilde{\omega}y \quad,\quad \pi^\pm _y = p_y \mp im\omega y - m\tilde{\omega}x . 
\end{equation}
 Note that so far, the system of equations is obtained without eliminating any component of the DKP spinor, in sharp contrast to Ref. \cite{PLA384:126706:2020}, which makes $e=0$ just to simplify the system of equations. Furthermore, making $e=0$ in (\ref{ec4})-(\ref{ec9}) our results are slightly different from those obtained in Refs. \cite{PLA384:126706:2020,PLA433:128030:2022}, due to their definition of the position vector as $\vec{r}=(-x,-y)$.

Before solving the system of equations (\ref{ec4})-(\ref{ec9}), it is very important to know which spinor components are physical components. This is achieved by applying operator $R^{\mu}$ (\ref{r2}) to the DKP spinor to obtain $R^{0}\psi = (0,0,-b,0,0,-e)^T$, $R^{1}\psi = (0,0,-a_2,0,0,d_1)^T$ and $R^{2}\psi = (0,0,a_1,0,0,d_2)^T$.  This result highlights two important points that deserve attention. The first point shows that all components of the six-spinor are physical components of the system, in this sense, it is an implausible claim to eliminate a priori any component of spinor because it would lead to a loss of information of the physical system. This point was addressed in a recent Comment \cite{PLA433:128030:2022}, where the authors proved that the spinor with $e=0$ leads to the trivial solution. The second point to consider is that from $R^{0}\psi$, $R^{1}\psi$, and $R^{2}\psi$, the physical components $(b,e)$, $(a_{2},d_{1})$ and ($a_{1},d_{2}$) are related in pairs (linearly dependent). By taking advantage of these relations between the physical components, it is possible to establish the equivalence between the set of equations (\ref{ec4})–(\ref{ec6}) and sets (\ref{ec7})–(\ref{ec9}) via the following relations: $a_{1}=-id_{2}=\Phi_{3}$, $a_{2}=id_{1}=-\Phi_{2}$, and $b=-ie=-i\Phi_{1}$. Consequently, the system of equations can be simplified to
\begin{equation}
m\Phi_1 = - \pi^-_y \Phi_2 + \pi^-_x\Phi_3, \label{ec12}
\end{equation}
\begin{equation}
m\Phi_2 = iE\Phi_3+ \pi^+_y \Phi_1, \label{ec13}
\end{equation}
\begin{equation}
m\Phi_3 = -iE\Phi_2- \pi^+_x \Phi_1. \label{ec14}
\end{equation}
\noindent It is straightforward to demonstrate that the above system of equations allows us to reconstruct a new DKPO equation following the same structure as (\ref{dkp_it}), considering only one of the two three-dimensional irreducible representations of the matrices $\beta^{\mu}$ and a three-component spinor $\Phi=(\Phi_{1},\Phi_{2},\Phi_{3})^{T}$, as predicted by \cite{CORSON1953}.. The $3\times 3$ matrices are given by \cite{PRB102:207101:2020}
\begin{equation}\label{r3x3}
\begin{split}
 &   \beta^0=\begin{pmatrix}
0 & 0 & 0\\
0 & 0 & i\\
0 & -i & 0
\end{pmatrix},\quad     \beta^1=\begin{pmatrix}
0 & 0 & -1\\
0 & 0 & 0\\
1 & 0 & 0
\end{pmatrix}, \quad    \\& \beta^2=\begin{pmatrix}
0 & 1 & 0\\
-1 & 0 & 0\\
0 & 0 & 0
\end{pmatrix}
\end{split}
\end{equation}
%\noindent Meanwhile, the time component $J^{0}$ becomes
%\begin{equation}\label{j0}
%J^{0}=\mathrm{Im}\left( \Phi_{3}^{\ast}\Phi_{2} \right)\,.
%\end{equation}
\noindent At this stage, we can conclude that in ($2+1$) dimensions one can use the representation (\ref{r3x3}) without unnecessary recurring to a six-dimensional representation, as used in \cite{JMP60:082302:2019,PLA384:126706:2020,chargui2023spin,PLA433:128030:2022}.

It should be noted that in Ref. \cite{chargui2023spin}, this construction of the DKP equation using irreducible representations was not performed. Instead, to decouple the $6\times 6$ system into two $3\times 3$ subproblems, a Hamiltonian-like structure was diagonalized through a unitary transformation. Nevertheless, within the rigorous prescription of DKP theory, this diagonalization should have been performed on the equation that represents the physical observable related to energy, that is, $\beta^0 H\Phi=\beta^0 E\Phi$, where $H$ is the Hamiltonian form of the system.

Returning our attention to the system of equations (\ref{ec12})-(\ref{ec14}), it is possible to combine the equations and after some algebraic manipulations we obtain a equation of motion for the $\Phi_{1}$ component
\begin{equation}
    \left(\vec{p}\,^2+\alpha^{2}r^{2}-2\gamma L_z-2\beta-E^2+m^2\right)\Phi_1=0    \label{total1}
\end{equation}
\noindent where
\begin{equation}
\alpha^2 = m^2(\omega^{2}+\tilde{\omega}^{2})+2Em\omega\tilde{\omega}, \label{ec15}
\end{equation}
\begin{equation}
\beta = E\tilde{\omega}+m \omega, \label{ec16}
\end{equation}
\begin{equation}
\gamma = E\omega +m\tilde{\omega}, \label{ec17}
\end{equation}

\noindent with $L_z=xp_y-yp_x$. The other physical components of the DKP spinor can be obtained by the expressions ($E\neq\pm m$)
\begin{equation}
\Phi_{2}= \frac{iE \pi^+_x  
-m \pi^+_y}{E^2-m^2}\Phi_{1}, \quad \Phi_{3} = \frac{iE \pi_y^+ +m \pi_x^+}{E^2-m^2}\Phi_{1} . \label{phi2}    
\end{equation}
\noindent In order to solve equation (\ref{total1}), we can use polar coordinates, and considering the usual decomposition
\begin{equation}
    \Phi_{1}(r,\varphi)=e^{il\varphi}\frac{\phi_{1}(r)}{\sqrt{r}}\,,
\end{equation}
\noindent with $l\in Z$, equation (\ref{total1}) becomes
\begin{equation}\label{eq_ef}
    \left(\frac{d^2}{d r^2}-\alpha^2 r^2-\frac{l^2-\frac{1}{4}}{r^2} +\kappa^2\right)\phi_1=0\,,
\end{equation}
\noindent where $\kappa^2=2\beta+E^2-m^2+2 l \gamma$.  Therefore, the solution of the DKP equation for spin-$1$ particles in the background of a ($2+1$)-dimensional DKPO plus an external magnetic field can be found by solving a Schr\"{o}dinger-like equation for the component of the DKP spinor $\phi_{1}$. The other components are obtained through of (\ref{phi2}). The equation (\ref{eq_ef}) has the form of equations (45) and (48) of the Refs. \cite{EPJC75:287:2015} and \cite{EPJC76:61:2016}, respectively. One finds that $\kappa^{2}>0$ and $\alpha^{2}>0$, and requiring $\phi_{1}$ to be square integrable, the solution for (\ref{eq_ef}) is precisely the well-known solution of the Schr\"{o}dinger equation for the harmonic oscillator. Note that the condition $\alpha^{2}>0$ implies that \begin{equation}\label{v1}
2E\omega\tilde{\omega}>-m\left( \omega^{2}+\tilde{\omega}^{2}\right)\,.
\end{equation}
\noindent Additionally, the condition $\kappa^{2}>0$ implies that $E>\epsilon_{+}$ and $E<\epsilon_{-}$, where
\begin{equation}\label{v2}
\epsilon_{\pm}=-\mu\pm\sqrt{\mu^{2}+m^{2}-\nu}\,,
\end{equation}
\noindent for $\mu^{2}+m^{2}-\nu\geq 0$ with $\mu=\omega l+\tilde{\omega}$ and $\nu=2m\left( \tilde{\omega}l+\omega \right)$, and for all $E\neq \pm m$ if $\mu^{2}+m^{2}-\nu< 0$.  The solution of (\ref{eq_ef}) for all $r$ and $\alpha>0$ can be written as 
\begin{equation}\label{ansatz2}
\phi_{1}(r)=r^{\vert l\vert+\frac{1}{2}}\mathrm{e}^{-\alpha r^{2}/2}f(r)\, .
\end{equation} 
\noindent Introducing the new variable  $ \rho = \alpha r^2$ and the following parameters 
\begin{equation}
  a= \frac{1}{2}\left(\vert l\vert+1-\frac{\kappa^2}{2\alpha}\right) ,  \quad b= \vert l\vert+1\ ,
\end{equation}
\noindent one finds that $f(\rho)$ can be expressed as a regular solution of the confluent hypergeometric equation, whose regular solution at $\rho=0$ is the Kummer function $f(\rho)=M\left(a,b,\rho\right)$.
\noindent By analyzing the asymptotic behavior of $f(\rho)$, it is possible to show that it diverges for $\rho \rightarrow \infty$. However, this can be remedied by demanding $a=-n_{r}$ and $b\neq-\tilde{n}$, where $n_r$ and  $\tilde{n}$ are a non-negative integer. In fact, $M(-n_r,b,\rho)$ with $b>0$ is proportional to the generalized Laguerre polynomial $L_{n_r}^{(b-1)}(\rho)$. Therefore, the solution for all $r$ can be written as
\begin{equation}\label{solg}
\phi_{1}(r)=Nr^{\vert l\vert+\frac{1}{2}}e^{-\alpha r^2/2}L^{(\vert l\vert)}_{n_r}(\alpha r^{2}),
\end{equation}
\noindent where $N$ is a normalization constant. Furthermore, the quantization condition $a=-n_r$ furnishes
\begin{flalign}
& 2\sqrt{ m^{2}(\omega^{2}+\tilde{\omega}^{2})+2Em\omega\tilde{\omega}}\left(2n_r+1+\vert l\vert\right) \notag \\
& = 2\omega(El+ m)+2\tilde{\omega}(E+m l)+E^2-m^2\,. \label{energ}
\end{flalign}
\noindent The solution of (\ref{energ}) determine the energy eigenvalues of our problem. This equation can be expressed as a fourth-degree algebraic equation in $E$. The solution of (\ref{energ}) can be obtained by searching energies that simultaneously satisfy the constraints (\ref{v1}), $E\neq\pm m$, $E>\epsilon_{+}$ and $E<\epsilon_{-}$ for $\mu^{2}+m^{2}-\nu\geq 0$, or simultaneously satisfy the constraints (\ref{v1}) and $E\neq\pm m$ for $\mu^{2}+m^{2}-\nu< 0$, as foreseen by the constraints on the parameters of the effective potential in (\ref{eq_ef}).

\section{Particular cases}
\label{section:pc}

Here, we present the solutions for the ($2+1$)-dimensional DKP oscillator and magnetic field.

\subsection{DKPO in ($2+1$) dimensions}
\label{section:pc:a}

For $\tilde{\omega}=0$, bound-state solutions are possible only for energies in the interval $E>\epsilon_{+}$ and $E<\epsilon_{-}$, where
\begin{equation}\label{v3}
\epsilon_{\pm}=-\omega l\pm\sqrt{(\omega l)^{2}+m^{2}-2m\omega}\,,
\end{equation}
\noindent for $(\omega l)^{2}+m^{2}-2m\omega\geq 0$ and for all $E$ if $(\omega l)^{2}+m^{2}-2m\omega<0$. In this case, the expression of the energy eigenvalues (\ref{energ}) reduces to
\begin{equation}
    E^2 +2\omega lE-2m\vert\omega\vert\left[2n_r+\vert l\vert+1-\mathrm{sgn}(\omega)\right]-m^2=0\,,
\end{equation}
\noindent which yields
\begin{equation}\label{ener_omega}
E_{n_r,l}=-\omega l\pm\sqrt{(|\omega l|+m)^{2}+ 2m\vert\omega\vert\left[2n_r+1-\mathrm{sgn}(\omega)\right]}\,.
\end{equation}
\noindent Furthermore,
\begin{equation}
\phi_{1}=Nr^{\vert l\vert+1/2}\mathrm{e}^{-m\vert\omega\vert r^{2}/2}L^{(\vert l\vert)}_{n_r}(m\vert\omega\vert r^{2})\,.
\end{equation}

The energy expression (\ref{ener_omega}) shows that for $l=0$, the discrete set of energies are symmetrical about $E=0$ and it is irrespective of the values of $n_r$ and $\omega$, but this symmetry does not hold for $l\neq 0$. It can be verified that the energy levels do not cross and that $\vert E\vert>m$. The expression (\ref{ener_omega}) also shows that one can obtain the energy spectrum for negative values of $l$ from the energy spectrum for positive values of $l$, and vice-versa. This result can be achieved by permuting simultaneously the signs of $l$ and $E$. This means that one can take advantage of this relation and without loss of generality one can focus our attention on positive values of $l$.

Now, we move on to study some symmetry related to $\omega$ from (\ref{ener_omega}). Defining $E_{+}$ and $n_{+}$ as $E$ and $n_{r}$ for $\omega>0$ and similarly, $E_{-}$ and $n_{-}$ as $E$ and $n_{r}$ for $\omega<0$, one finds that
\begin{equation}\label{vin_omega}
\vert E_{+}+\vert\omega\vert l\vert=\vert E_{-}-\vert\omega\vert l\vert\,,
\end{equation}
\noindent if and only if $n_{+}=n_{-}+1$. In particular, there is no state with $n_{r}=0$ for $\omega>0$, and $\vert E_{+}\vert=\vert E_{-}\vert$ when $l=0$.

In the non-relativistic limit, $E=m+\varepsilon$ with $m\gg\vert \varepsilon\vert$ and $m\gg\vert\omega\vert$, the Eq.~(\ref{ener_omega}) becomes
\begin{equation}\label{lnr_ener_omega}
\varepsilon\simeq \vert\omega\vert\left\{2n_r+1-\mathrm{sgn}(\omega)+\vert l\vert\left[1-\mathrm{sgn}(\omega)\mathrm{sgn}(l)\right]\right\}\,.
\end{equation}

\begin{figure}[ht]
\begin{subfigure}{0.39\textwidth}
\centering
\includegraphics[width=\textwidth]{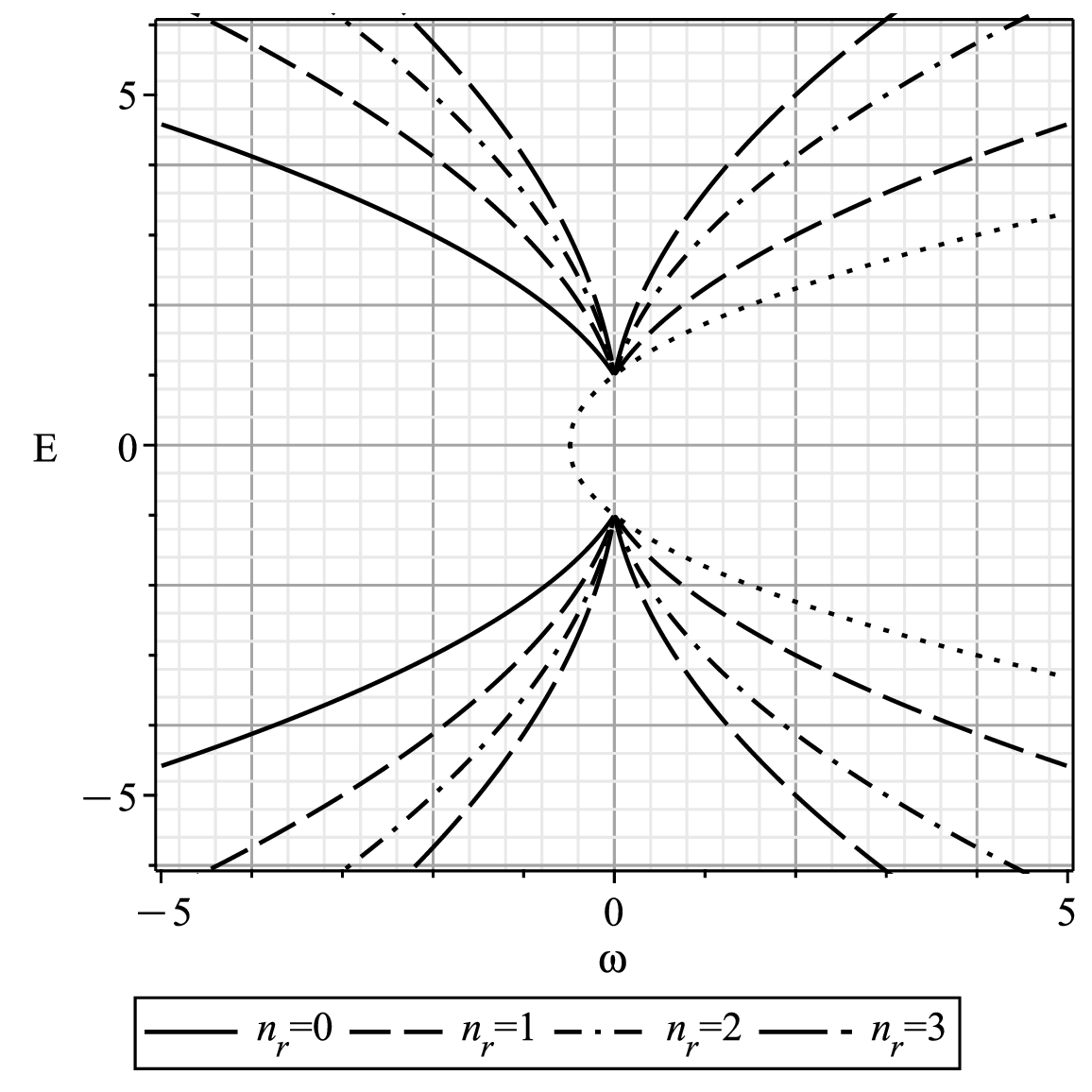}
\caption{$l=0$}
\label{fig:EvsOmega_l0}
\end{subfigure}
\begin{subfigure}{0.39\textwidth}
\centering
\includegraphics[width=\textwidth]{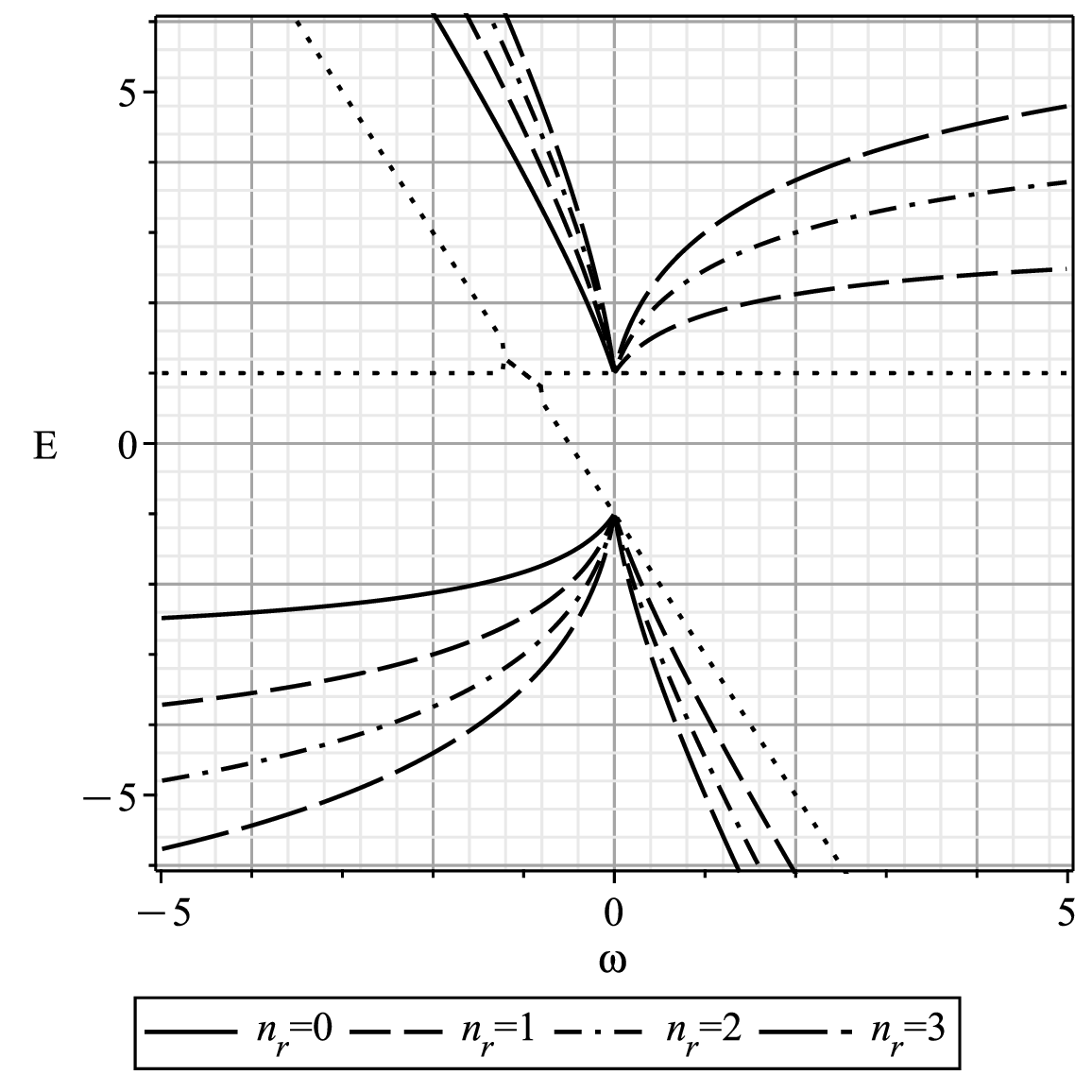}
\caption{$l=1$}
\label{fig:EvsOmega_l1}
\end{subfigure}
\caption{Plots of the energy as a function of $\omega$ and different values of $n$ and $l$, with $m=1$. The dotted line represents the constraints $\epsilon_{+}$ and $\epsilon_{-}$.}
\label{fig:EvsOmega}
\end{figure}

Figure \ref{fig:EvsOmega} illustrates the profiles of the energy as a function of $\omega$ for different values of $n$ and $l$, with $m=1$. We consider the four first principal quantum numbers and two different values of $l$. From Figs. \ref{fig:EvsOmega_l0} and \ref{fig:EvsOmega_l1}, one sees that the energies for $n_{r}=0$ and $\omega>0$ do not belong to the spectrum of allowed energies. Also, it is noticeable from these figures that all the energy levels emerge from the positive (negative)-energy continuum and that for positive (negative) energy spectrum one finds that the lowest quantum number $n_{r}$ with a fixed value of $l$ correspond to the lowest (highest) energy level, as it should be for particle (antiparticle) energy levels. For $l=0$ (Fig. \ref{fig:EvsOmega_l0}), one notes that the discrete set of energies are symmetrical about $E=0$, as expected.

\begin{figure}[ht]
\begin{subfigure}{0.39\textwidth}
\centering
\includegraphics[width=\textwidth]{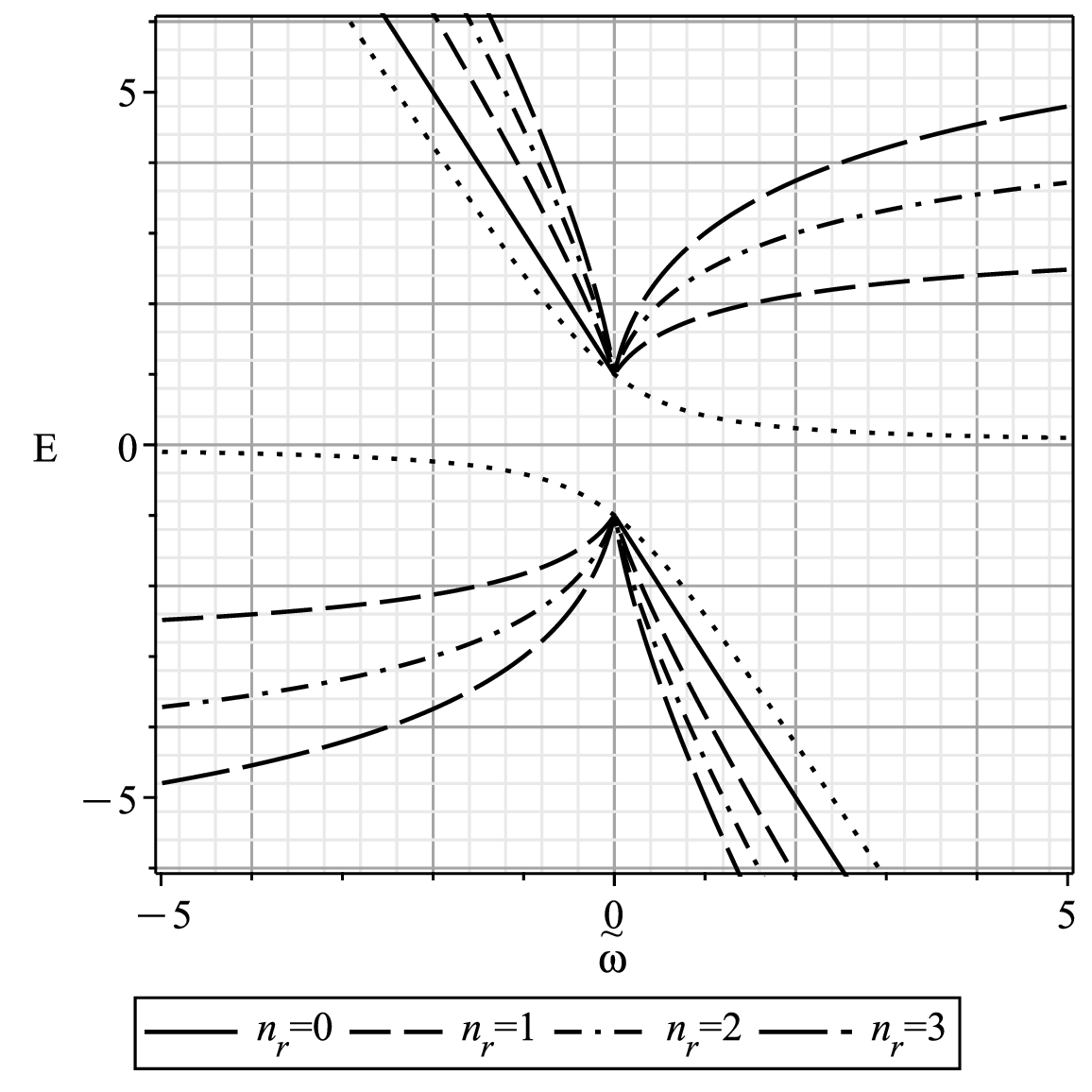}
\caption{$l=0$}
\label{fig:EvsOmegatilde_l0}
\end{subfigure}
\begin{subfigure}{0.39\textwidth}
\centering
\includegraphics[width=\textwidth]{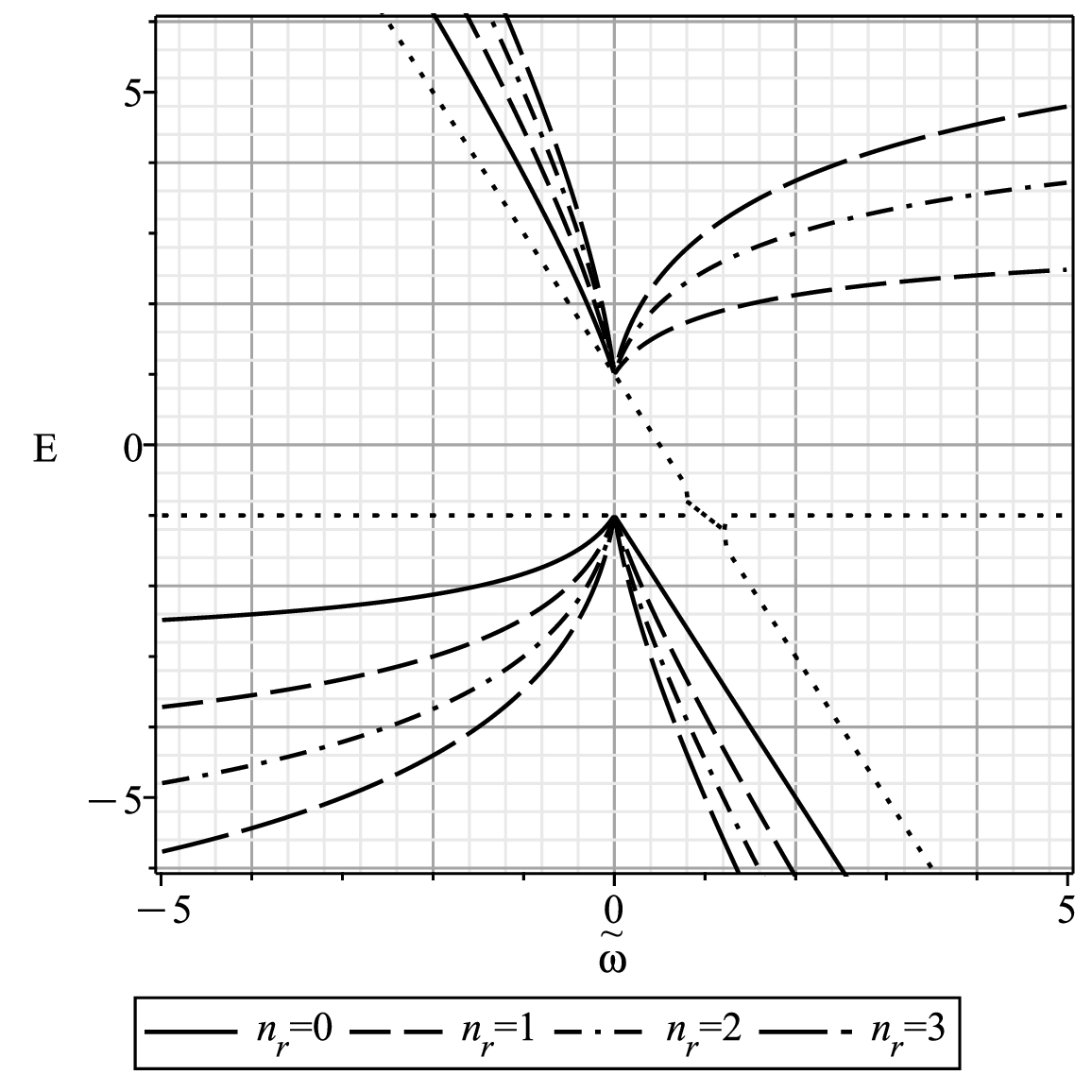}
\caption{$l=1$}
\label{fig:EvsOmegatilde_l1}
\end{subfigure}
\caption{Plots of the energy as a function of $\tilde{\omega}$ and different values of $n$ and $l$, with $m=1$. The dotted line represents the constraints $\epsilon_{+}$ and $\epsilon_{-}$.}
\label{fig:EvsOmegatilde}
\end{figure}
\subsection{Uniform magnetic field in ($2+1$) dimensions}
\label{section:pc:b}
For $\omega=0$, bound-state solutions are possible only for energies in the interval $E>\epsilon_{+}$ and $E<\epsilon_{-}$, where
\begin{equation}\label{v4}
\epsilon_{\pm}=-\tilde{\omega}\pm\sqrt{\tilde{\omega}^{2}+m^{2}-2m\tilde{\omega}l}\,,
\end{equation}
\noindent for $\tilde{\omega}^{2}+m^{2}-2m\tilde{\omega}l\geq 0$ and for all $E$ if $\tilde{\omega}^{2}+m^{2}-2m\tilde{\omega}l<0$. In this case, the expression of the energy eigenvalues (\ref{energ}) reduces to
\begin{equation}
    E^2 +2\tilde{\omega} E-2m\vert\tilde{\omega}\vert\left\{2n_r+1+\vert l\vert\left[1-\mathrm{sgn}(\tilde{\omega})\mathrm{sgn}(l)\right]\right\}-m^2=0\,,
\end{equation}
\noindent which yields
\begin{equation}
\begin{aligned}
E_{n_r,l} = & -\tilde{\omega} \pm \\
  & \sqrt{(|\tilde{\omega}|+ m)^2+ 2m\vert\tilde{\omega}\vert\left\{2n_r+\vert l\vert\left[1-\mathrm{sgn}(\tilde{\omega})\mathrm{sgn}(l)\right]\right\}}\,.
\end{aligned}
\label{ener_omegatilde}
\end{equation}
\noindent Furthermore,
\begin{equation}
\phi_{1}=Nr^{\vert l\vert+1/2}\mathrm{e}^{-m\vert\tilde{\omega}\vert r^{2}/2}L^{(\vert l\vert)}_{n_r}(m\vert\tilde{\omega}\vert r^{2})\,.
\end{equation}
The energy expression (\ref{ener_omegatilde}) shows that the energies are never symmetrical about $E=0$. The energy gap between positive and negative energy levels is $2m$ at least, and those levels never cross. The expression (\ref{ener_omegatilde}) also shows that one can obtain the energy spectrum for negative values of $l$ from the energy spectrum for positive values of $l$, and vice-versa. This result can be achieved by permuting simultaneously the signs of $l$, $\tilde{\omega}$ and $E$. In particular, for $l=0$ the energies are symmetrical under the changes $\tilde{\omega}\rightarrow-\tilde{\omega}$ and $E\rightarrow-E$. As in the previous subsection, this means that one can take advantage of this relation and without loss of generality one can focus attention on positive values of $l$.

At this stage, we move on to study some symmetry related to $\tilde{\omega}$ from (\ref{ener_omegatilde}). Defining $E_{+}$ and $n_{+}$ as $E$ and $n_{r}$ for $\tilde{\omega}>0$ and similarly, $E_{-}$ and $n_{-}$ as $E$ and $n_{r}$ for $\tilde{\omega}<0$, one finds that
\begin{equation}\label{vin_omegatilde}
\vert E_{+}+\vert\tilde{\omega}\vert \vert=\vert E_{-}-\vert\tilde{\omega}\vert \vert\,,
\end{equation}
\noindent if and only if $n_{+}=n_{-}+l$.

In the non-relativistic limit, $E=m+\varepsilon$ with $m\gg\vert\varepsilon\vert$ and $m\gg\vert\tilde{\omega}\vert$, the Eq.~(\ref{ener_omegatilde}) becomes
\begin{equation}\label{lnr_ener_omega}
\varepsilon\simeq \vert\tilde{\omega}\vert\left\{2n_r+1-\mathrm{sgn}(\tilde{\omega})+\vert l\vert\left[1-\mathrm{sgn}(\tilde{\omega})\mathrm{sgn}(l)\right]\right\}\,.
\end{equation}

Figure \ref{fig:EvsOmegatilde} illustrates the profiles of the energy as a function of $\tilde{\omega}$ for different values of $n$ and $l$, with $m=1$. We consider the four first principal quantum numbers and two different values of $l$. From Figs. \ref{fig:EvsOmega_l0} and \ref{fig:EvsOmega_l1}, one sees that the positive (negative) energies are to be identified with particle (antiparticle) levels, following the same interpretation as in the previous subsection. Note that for $l=0$ (Fig. \ref{fig:EvsOmegatilde_l0}) and $n_{r}=0$, the energy for particles (antiparticles) with $\tilde{\omega}>0$ ($\tilde{\omega}<0$) does not belong to the spectrum of allowed energies. A similar behavior is found in Fig. \ref{fig:EvsOmegatilde_l1} ($l=1$), but in this case only the energy for particles for $n_{r}=0$ with $\tilde{\omega}>0$ does not belong to the spectrum of allowed energies.

Next, we present a theoretical prescription that utilizes the 2D-DKP theory in the context of condensed matter systems, focusing on its ability to calculate specific electronic properties in Lieb lattices. This framework followed an approach similar to that used for graphene.
%\cleardoublepage
%%%%%%%%%%%%%%%%%%Spin-1 quasiparticles in Lieb lattice: A quantum electrodynamical approach
\section{Two-dimensional DKP prescription and the connection with condensed matter physics}
\begin{figure}[h]
\centering
\includegraphics[width=4.0cm]{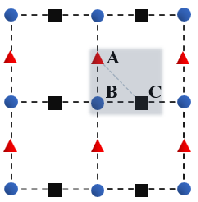}
\caption{Lieb lattice}
\label{fig:Lieb}
\end{figure}

The Lieb lattice, illustrated in Figure \ref{fig:Lieb}, is a 2D face-centered square lattice and characterized by three lattice sites (A, B, C) per unit cell. Similar to graphene, in a Lieb lattice, electrons are able to move between nearest neighbors. Specifically, electrons on a type B atom can hope to the nearest A or C atom. This behavior is represented by the tight-binding Hamiltonian $\mathcal{H}_0=t\sum_{<ij>}(b^{\dagger}_{i}a_{j}+b^{\dagger}_{i}c_{j}+h.c.)$, where  the sum is over nearest neighbors $<$$ij$$>$,  with hopping amplitude $t$.

Nevertheless,  hopping between A and C atoms is possible in specific scenarios, such as when considering circulating current states (Varma phase) \cite{PRB97:075135:2018} or when examining the effects of spin-orbit coupling. For both cases, we add a term of the form $a^{\dagger}_ic_j$ along with the Hamiltonian $\mathcal{H}_0$, resulting in the emergence of a bandgap energy. Following a procedure similar for graphene, it is possible to demonstrate that the dynamics of the charge carriers on the Lieb lattice in the low-energy regime exhibit a relativistic-like behavior \cite{PRB102:207101:2020}, as described by the two-dimensional DKP-like Hamiltonian:
\begin{equation}
\hat{H}_{DKP}\Phi =E\Phi ,  \label{67}
\end{equation}%
\begin{equation}
\hat{H}_{DKP}=v_F\left[ \beta ^{0},\beta ^{1}\right] \hat{p}^{1}+v_F\left[ \beta ^{0},\beta
^{2}\right] \hat{p}^{2}+m\beta ^{0},  \label{68}
\end{equation}
where $v_F$ is the relative Fermi velocity, and $m$ represents the bandgap energy. Here, the same $\beta$-matrices defined in (\ref{r3x3}) emerge naturally owing to the three-band structure of the Lieb lattice. Therefore, the three-component spinor $\Phi=(\Phi_a, \Phi_b, \Phi_c)^T$ correctly describes spin-1 quasiparticles, similar to that studied in this Remark. Nevertheless, it is necessary to establish the equivalence between the Hamiltonian form (\ref{68}) and the equation of motion (\ref{ec-dkp}), which is achieved using the constraint equation $v_F\beta ^{i}\beta ^{0}\beta ^{0}p_{i}\Phi =m\left( 1-\beta ^{0}\beta^{0}\right) \Phi$ \cite{EPJC75:287:2015,PRB102:207101:2020}. 

In this scenario, it is possible to identify the potential physical reality of the system addressed in this study. For example, when computing the electronic properties of nanostructures based on the Lieb lattice, such as electrical conductivity, it is advantageous to extend the single-particle prescription given in (\ref{68}) to the framework of one-loop corrections in quantum electrodynamics for DKP spin-1 quasiparticles \cite{PRB97:075135:2018,PRB102:207101:2020}. Thus, the corresponding effective action $S_{ \text{eff}}(A)=i\text{ln Det}[\tilde{\beta} ^{\mu}(p_\mu-eA_\mu)-m]$ yields the following polarization tensor
\begin{equation}
i\Pi ^{\mu \nu }\left( p\right)=e^{2}\int \frac{d^{3}k}{%
(2\pi )^{3}}\text{Tr}\left[ \tilde{\beta} ^{\mu }G_{\Phi }(k-p)\tilde{\beta} ^{\nu }G_{\Phi
}(k)\right], \label{69}
\end{equation}
where $\tilde{\beta^0}=\beta^0$, $\tilde{\beta^i}=v_F\beta^i$ and
\begin{equation*}
G_{\Phi }(k)=i\frac{1}{\tilde{\beta} ^{\mu }k_{\mu }-m}=\frac{i}{m}\left[ \frac{%
\slashed{k}(\slashed{k}+m)}{\tilde{k}^{2}-m^{2}}-1\right],
\end{equation*}
is the DKP free (Feynman) propagator. For illustrative purposes, we calculate some elements of (\ref{69}) using the properties of the trace of the matrices $\beta$ defined in \cite{PRB102:207101:2020} along with the standard techniques employed for solving Feynman integrals. Thus, for example, we have
\begin{equation}
    i \Pi^{ij}(p)= \frac{ie^2}{2\pi} \left[\frac{v^2_Fp_i p_j}{|m|}  \Pi_{even}(p)-i\epsilon^{ij0}p_0 \, \Pi_{odd} (p) \right], \label{70}
\end{equation}
where $i\neq j$ and $\Pi_{even}$ and $\Pi_{odd}$ are dimensionless parameters
\begin{eqnarray*}
    \Pi_{even} (p) &=& \int^1_0 \frac{x(1-x)\left(1-\frac{1}{4}\frac{\tilde{p}^2}{m^2} \right)}{\sqrt{1-x(1-x)(\tilde{p}^2/m^2)}} dx,\\
    \Pi_{odd}(p) &=& \text{sign} (m) \int^1_0 \sqrt{1-x(1-x)(\tilde{p}^2/m^2)} dx.
\end{eqnarray*}
where $\text{sign} (m)$ represents the sign of the surface bandgap. In the large surface bandgap limit ($|m|\rightarrow\infty$), the term proportional to $\Pi_{even}$ vanishes, whereas $\Pi_{odd}\rightarrow\ \text{sign} (m)$ corresponds to a topological term in the effective action $S_{ \text{eff}}(A)$. In this limit, the electrical conductivity is calculate through the Kubo formula, as performed in \cite{PRB102:207101:2020}. Moreover, the expression in (\ref{70}) along with $i \Pi^{00}(p)= \frac{ie^2}{2\pi}\frac{\boldsymbol{p}^2}{|m|}\Pi_{even} (p)$, would also facilitate the examination of the Casimir interaction between topological insulators based on Lieb lattice, similar to the approach taken in \cite{Liucasimir} for graphene.
%%%%%%%%%%%%%%%%%%
\section{Conclusions}
We have re-examined the problem of spin-$1$ particles in the background of a $(2 + 1)$-dimensional DKPO in the presence of a uniform magnetic field. Following the appropriate procedure for the spin-$1$ sector of the Duffin-Kemmer-Petiau theory, we use a particular $6\times 6$ matrix representation for the matrices $\beta^{\mu}$ \cite{JMP60:082302:2019,PLA384:126706:2020,chargui2023spin,PLA433:128030:2022} and we show that it can be reduced to two equivalent irreducible representations of $3\times 3$ matrices. Unlike the approach used in \cite{chargui2023spin}, the system of equations (\ref{ec12}), (\ref{ec13}), and (\ref{ec14}) allowed for the reconstruction of the DKP equation using this $3\times 3$ irreducible representation (as given in (\ref{r3x3})). We found a coupled system of equations for the physical components of DKP spinor, given by the Eqs. (\ref{ec12}), (\ref{ec13}) and (\ref{ec14}). Without making any arbitrary exclusion of DKP spinor components, as done in \cite{PLA384:126706:2020}, we show that the second order differential equation for the $\phi_{1}$ component can be mapped into a confluent hypergeometric differential equation in cylindrical coordinates. In this way, the DKP spinor and energy spectrum were exactly obtained. Additionally, we have studied the cases of DKPO (4.1) and uniform magnetic field (4.2) as particular cases. In both cases, a rigorous analysis was conducted on the allowed values for each energy, considering the existing constraints among the system parameters. Finally, the successful implementation of the DKP equation in (2+1)-dimensional space-time, utilizing a $3\times 3$ irreducible representation of beta-matrices, within condensed matter systems, specifically, in the Lieb lattice as detailed in \cite{PRB97:075135:2018,PRB102:207101:2020}. The extension of the single-particle prescription (\ref{68}), allowing for one-loop corrections in quantum electrodynamics for DKP spin-1 quasiparticles, would enable us to determine electronic properties, such as electrical conductivity, as well as to examine the Casimir interaction between topological insulators based on the Lieb lattice.
\acknowledgments
This work was supported in part by means of funds provided by CNPq, Brazil, Grants No. 09126/2019-3 and 311925/2020-0, FAPEMA and CAPES - Finance code 001.
\bibliographystyle{spphys}       % APS-like style for physics
%\bibliography{mybibfile2020}

\end{document}